# Meeting Global Cooling Demand with Photovoltaics during the 21st Century

Hannu S. Laine[a,b], Jyri Salpakari[c], Erin E. Looney[a], Hele Savin[b], Ian Marius Peters[a] and Tonio Buonassisi[a]

Space conditioning, and cooling in particular, is a key factor in human productivity and well-being across the globe. During the 21st century, global cooling demand is expected to grow significantly due to the increase in wealth and population in sunny nations across the globe and the advance of global warming. The same locations that see high demand for cooling are also ideal for electricity generation via photovoltaics (PV). Despite the apparent synergy between cooling demand and PV generation, the potential of the cooling sector to sustain PV generation has not been assessed on a global scale. Here, we perform a global assessment of increased PV electricity adoption enabled by the residential cooling sector during the 21st century. Already today, utilizing PV production for cooling could facilitate an additional installed PV capacity of approximately 540 GW, more than the global PV capacity of today. Using established scenarios of population and income growth, as well as accounting for future global warming, we further project that the global residential cooling sector could sustain an added PV capacity between 20-200 GW each year for most of the 21st century, on par with the current global manufacturing capacity of 100 GW. Furthermore, we find that without storage, PV could directly power approximately 50% of cooling demand, and that this fraction is set to increase from 49% to 56% during the 21st century, as cooling demand grows in locations where PV and cooling have a higher synergy. With this geographic shift in demand, the potential of distributed storage also grows. We simulate that with a 1 $m^3$ water-based latent thermal storage per household, the fraction of cooling demand met with PV would increase from 55% to 70% during the century. These results show that the synergy between cooling and PV is notable and could significantly accelerate the growth of the global PV industry.

**Broader context**

This article analyses the intersection of two dominant energy trends of the 21st century: 1) The impetus to decarbonize the energy sector to prevent dangerous climate change and 2) The rapidly increasing global cooling demand due to increased economic prosperity in tropical countries. Specifically, we investigate whether the several billions of air-conditioning devices expected to come online this century could be powered by clean photovoltaic (PV) electricity, avoiding the need for additional carbon-based electricity. To answer that question, we pool nearly a terabyte of high-resolution weather and population data as well as socio-economic, global warming and energy efficiency scenarios and input them into globally validated models for cooling demand, PV electricity generation and thermal storage. We show that the potential added PV capacity is significant, more than the global PV capacity today and spanning several TW's by the end of the century, accelerating the PV industry in the process. We show that the geographic shift of cooling demand closer to the equator significantly improves the synergy of cooling, PV and small-scale distributed storage. These findings provide quantitative evidence of the large scope and high degree of synergy between cooling and PV to investors, policy-makers and researchers.

## Introduction

Maintaining a comfortable ambient temperature is a central factor in human productivity and spread across the globe. The availability of space heating and cooling impacts all aspects of our lives from sleep quality[1,2], cognitive capacity[3], temperament[4] up to the culture[5,6] and social fabric[7] of our societies.

During the 20th century, the energy demand of space conditioning was mostly driven by increased use of heating in wealthy colder nations[8]. In the 21st century, tropical nations are gaining wealth, bringing much needed air conditioning to billions of people and acting as a key driver for human prosperity. Air-conditioning usage is expected to further increase due to global warming[9], possibly becoming a strict requirement for human survival in the hottest locations[10].

The global electricity usage for cooling in 2016 was 2000 TWh, or 18.5% of annual electricity consumption in buildings[11]. Moreover, the residential cooling sector is expected to grow by more than 10-fold during the 21st century[12,13]. Locally, cooling demand often comprises over 50% of peak electricity consumption for cities in both developed and developing countries[14], a trend which is further aggravated by climate change. This is a massive, highly variable, added electricity load in the next century, and with the urgent need to decrease anthropogenic carbon emissions, this electricity must be mainly produced via with low-carbon energy sources to protect the planet from further global warming[15].

Within this challenge of increasing air conditioning with clean energy lies an opportunity: the possibility to use the very source of the heat, the sun, to power added air conditioning devices. In principle, cooling is most needed during the hottest hours of the day, when solar insolation, and hence, photovoltaic (PV) electricity production, is at its highest. High temperatures and high solar insolation also correlate geographically, with sunny countries experiencing warmer climates, and thus, needing more air-conditioning.

Due to the high potential of air-conditioning to enable significant grid integration of PV, the cooling sector has the potential to influence decarbonisation of electricity production, global energy policy and PV technology roadmaps[16-18]. Quantifying the potential synergy of AC and PV helps both the AC and PV industry align their marketing strategies and helps governments to plan possible policy actions. Even though the

[a.] Massachusetts Institute of Technology, Cambridge, MA 02139, USA.
[b.] Department of Electronics and Nanoengineering, Aalto University, 02150 Espoo Finland.
[c.] New Energy Technologies Group, Department of Applied Physics, Aalto University, 02150 Espoo, Finland.

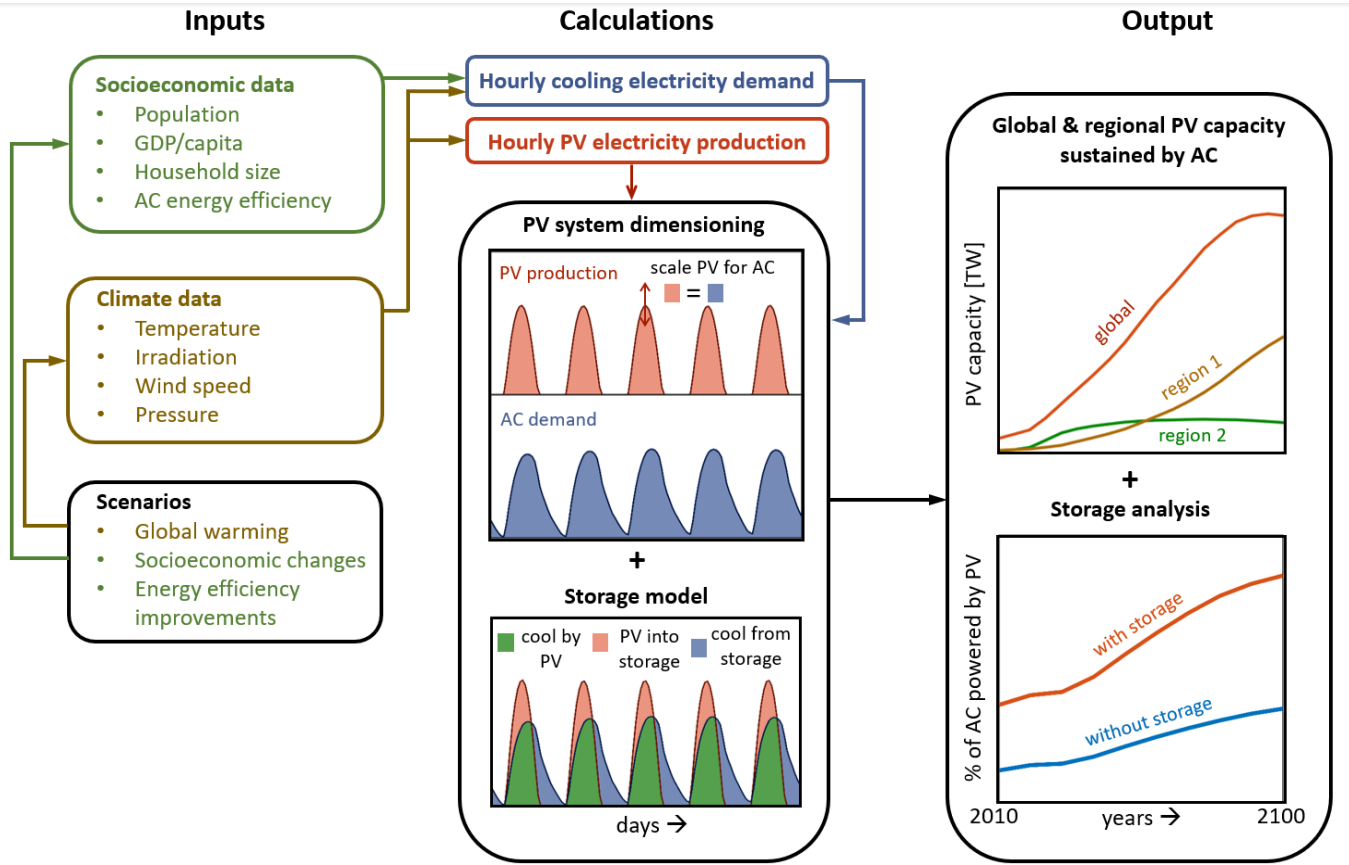

**Fig. 1** A schematic of the model used in the paper

value of the cooling sector in increasing PV penetration at large scale has been recognized[19,20], and detailed studies have been conducted at a single building or community scale[21-30], the potential synergy of PV and cooling has not been quantified globally.

Here, we estimate the electricity demand of the global residential cooling sector today and throughout the 21st century and calculate the corresponding PV capacity that the sector could sustain. Additionally, we investigate the different time scales in the temporal mismatch of cooling demand and PV electricity production and to what extent distributed thermal storage could mitigate the mismatch.

## Models and data

We use a socioeconomic model to predict AC electricity consumption globally[12], accounting for the tendency of people to obtain and utilize cooling devices depending on their wealth and local weather. With the help of a PV system simulator[31], we then calculate the size of a PV system that would match the cooling electricity consumption on a yearly basis at each modelled location. We then evaluate the fraction of diurnal and seasonal storage needed to mitigate the different dynamics of cooling demand and PV electricity production, and how these needs change throughout the century across the globe. Lastly, we employ a thermal storage model for a realistic storage scenario. The model schematic is described in Fig. 1.

### Socioeconomic and climate scenarios and datasets

To accurately match population density with local weather, we draw on a global population-grid with a resolution of 30''×30'' (approximately 1 km at the equator)[32] and match these data with an hourly time-series of a 0.4°×0.4° (approximately 40 km at equator) resolution grid of global climate data[33]. Hourly resolution is sufficient, because building heat capacity facilitates intra-hourly matching of PV production to air-conditioning demand[34]. The weather data is from years 1985-2005. To estimate long-term trends, we average each hour of the year across the 21 years of data and increase the temperature each year due to global warming[35].

For country-level scenarios on the growth of population and gross domestic product (GDP), we use the Shared

**Table 1** Explored socioeconomic scenarios

| Scenario | Population/ billion | | GDP per capita/ 1000 $US* | |
|---|---|---|---|---|
| | 2050 | 2100 | 2050 | 2100 |
| SSP2 | 9.1 | 9.0 | 25.1 | 59.6 |
| SSP3 | 10.0 | 12.7 | 17.8 | 21.9 |
| SSP5 | 8.5 | 7.4 | 42.2 | 137.7 |

*all GDP per capita values in this paper are listed in purchasing power parity (PPP) adjusted US 2005 dollars.

Socioeconomic Pathways (SSP)[36], which have been developed by the international climate change modelling community for modelling climate change impacts and mitigation during the 21st century. Of the 5 SSP's available, we choose SSP2 as the reference case, and explore SSP3 and SSP5 as examples of slow and fast socioeconomic development. These scenarios provide the lowest and highest predicted PV deployment by mid-century, respectively, among the five pathways. The population and GDP per capita development in these scenarios are listed in Table 1. For reference, in 2010, world population was approximately 6.8 billion and GDP per capita 10 000 $US.

For climate change, we employ Representative Concentration Pathways (RCP) developed by the Intergovernmental Panel on Climate Change (IPCC)[37]. Our reference case includes 2.5°C warming by the end of the century ("RCP4.5"), and we similarly explore a rapid mitigation scenario with 1.8°C ("RCP3") and a business-as-usual scenario with 4.5°C ("RCP8.5") of warming to elucidate the relative impact of global warming.

For household size data, we take country-level values provided by the UN until 2030[38], and assume that after 2030, the household size correlates with the projected total fertility rate[39]. We assume the average coefficient of performance of air-conditioners to increase from 2.4 in 2000 to 4.39 in 2100, as projected in Ref. 40.

**Model for residential cooling electricity demand**
To model the evolution of cooling electricity demand around the globe, we use the approach developed by Isaac and van Vuuren[12]. We focus here on the residential cooling sector, as its growth drivers are better quantified than those of the commercial sector. Globally, the commercial cooling sector consumes nearly as much electricity as the residential sector today, although it is expected to grow slower[11].

We calculate the residential cooling demand for every year, accounting for projected changes in socioeconomic and climate parameters, as well as improved energy efficiency of cooling devices. The model inputs include population, household size and GDP data, as well as local temperature time-series. For a given population, the annual electricity used for cooling, $E_{pop}^a$, is a product of five variables:

$$E_{\text{pop}}^a = N \times A \times S_{\max} \times E_{\text{h}} \times \eta, \quad (1)$$

1) *The number of households within the population, $N$*: Population divided by the average number of people in a household.
2) *Availability, $A$*: The fraction of households which can afford air-conditioning.
3) *Climate maximum saturation, $S_{\max}$*: The fraction of households, which would acquire air-conditioning if they could afford it, given a certain climate. The product $A \times S_{\max}$ determines the total fraction of households with air-conditioning.
4) *Annual household electricity consumption, $E_{\text{h}}$*: The average annual electricity used for cooling by each household with air-conditioning.
5) *Energy efficiency parameter, $\eta$*: which describes the energy efficiency of air-conditioners improving over time.

The climate-related parameter influencing these variables is the number of cooling degree days, $CDD$:

$$\text{If } T_d(t) > T_{\text{base}}, \quad CDD(t) = T_d(t) - T_{\text{base}} \quad (2)$$
$$\text{else}, \quad CDD(t) = 0,$$

where $T_d$ is the daily mean temperature and the base temperature $T_{base}$ is set to 18°C[12,41]. The base temperature implicitly accounts for unintentional sources of heating, such as inhabitants and appliances. The climate maximum saturation depends on annual cooling degree days, $CDD_a$, exponentially[12,42]:

$$S_{\max} = 1 - 0.949 \times \exp(-0.00187 \times CDD_a). \quad (3)$$

For reference, $S_{\max}$, reaches 90% saturation around 1200 $CDD_a$ (=average of 10°C cooling for 120 days), a climate slightly hotter than, for example, Washington, D.C. (1140 $CDD_a$). The availability $A$, on the other hand, depends on wealth of the population, measured in GDP per capita, according to a logistic function fitted in Ref. 12 based on the empirical data in Refs. 42-47:

$$A = 1/(1 + \exp(-0.304/1000 \times GDP/cap + 4.152)). \quad (4)$$

For reference, 90% availability is reached at around 43 000 $ per capita, corresponding to GDP per capita in today's United Kingdom. Lastly, $E_h$ is dependent on climate and income (wealthier people tend to have larger houses):

$$E_{\text{h}} = CDD_a \times (0.865 \times \ln GDP/cap - 5.825). \quad (5)$$

The energy efficiency parameter $\eta$, is scaled such that the average coefficient of performance of air-conditioners increases from 2.4 in 2000 to 4.39 in 2100, as projected in Ref. 40.

**Hourly dynamics of cooling demand.** When estimating the synergy of cooling and PV, a high time resolution is important for both the cooling and PV energy yield time series. Hourly resolution is typically deemed sufficient, because intra-hour load matching is facilitated by the thermal inertia of buildings[34]. To derive hourly time-series from the annual electricity usage $E_{\text{pop}}^a$, we follow the approach in Ref. 41 by accounting for the hourly distribution of cooling demand *via* cooling degree hours *CDH*, and the hourly distribution of the coefficient of performance, *COP*, of the cooling devices. *COP* describes how many units of thermal energy can be transferred with a single unit of electrical energy. *COP* is typically >1 and grows higher the smaller the temperature difference between the building interior and ambient temperature is. We model the air-conditioners as Carnot heat pump cycles[30], with the Carnot efficiency $\eta_{Ca}$ describing cycle non-ideality, and the temperature differences $\delta T_1 = \delta T_2 = 5°C$ describing heat exchangers:

$$COP(t) = \eta_{Ca} \times \frac{T(t) - \delta T_1}{T_{base} + \delta T_2 - (T(t) - \delta T_1)}, \quad (6)$$

where $\eta_{Ca}$ improves from 0.16 in 2000 to 0.29 in 2100, which corresponds to similar $COP$ development as predicted in Ref. 40. With the hourly-resolved $COP$ and $CDH$, the hourly time-series for cooling electricity consumption, $E_{pop}^h$ is then reached via:

$$E_{pop}^h(t) = E_{pop}^a \times \frac{CDH(t)}{\sum_{i=1}^{N_h} CDH(i)} \left(\frac{COP(t)}{\sum_{i=1}^{N_h} COP(i)}\right)^{-1}, \quad (7)$$

where $N_h$ is the number of hours in a year and $E_{pop}^a$ is calculated with Eq. (1).

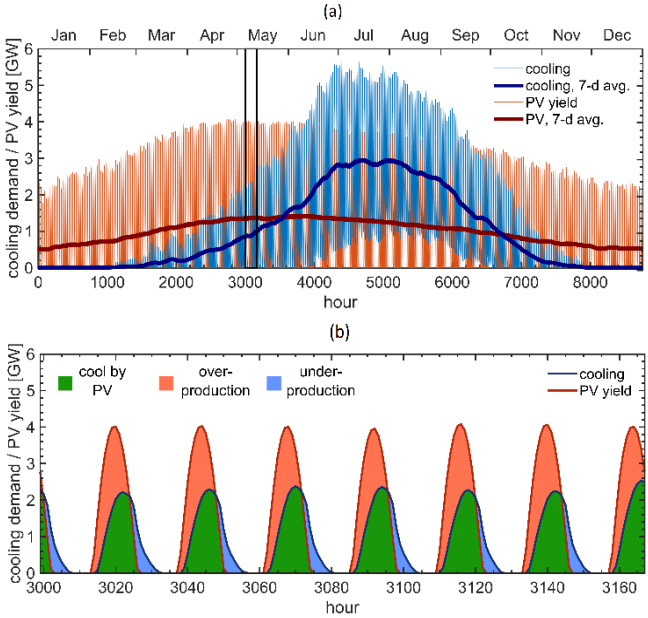

**Fig. 2** Modelled AC demand and PV electricity yield for (a) year 2010 (b) 2nd week of May, 2010 for Maricopa County, AZ, which is host to the state capital, Phoenix. The population of the county was approximately 3.8 million at the time. The black box in Fig. (a) highlights the week chosen for Fig. (b). Our model performs a similar PV system dimensioning for each geographical location and for each modelled year.

**PV electricity yield model**
We use the Matlab-based open-source PV Lib toolbox developed by Sandia National Laboratories[31] to model PV electricity production. For each geographical location, we calculate a normalized energy yield time series (hourly kWh produced / W) and scale the system with the algorithm described in the next section.
The hourly energy yield is determined by the available sunlight, sun's location, temperature, wind speed and pressure. As inputs, we use global horizontal irradiance (GHI), temperature, wind speed and pressure from the same climate dataset[33] that is used to calculate the cooling electricity demand. We calculate the direct normal irradiance and the diffuse horizontal irradiance using the DISC model[48]. We assume a fixed system with a tilt angle equal to the latitude of the location, with arrays facing south in the Northern hemisphere and north in the Southern hemisphere. The Sandia PV Array Performance Model accounts explicitly for losses in module performance at higher temperatures. Other loss mechanisms, such as module degradation[49], current mismatch of modules[50], DC-AC conversion[51], soiling[52], partial shading of modules[53], and maximum power point tracking[53] are accounted for by decreasing the hourly energy yield by 10%[54] each hour. For modelling the cell temperature and air-mass, we use the default values in the Sandia Module Database version dated September 29, 2012[31].

**PV system dimensioning algorithm**
After reaching an hourly estimate of AC electricity use and a normalized energy yield for each census location around the globe, we dimension the PV system that is used to power the AC. We scale the PV capacity for each census location such that the yearly PV production matches the yearly electricity used for cooling:

$$\sum_{t=1}^{N_h} E_{pop}^h(t) = \sum_{t=1}^{N_h} E_{PV}(t), \quad (8)$$

where $E_{PV}(t)$ is the hourly electricity production of the PV system, and the summation is performed over a full year. The PV capacity is increased every year as more households acquire cooling devices and the average household cooling demand grows due to socioeconomic and climatic changes.

**PV system dimensioning example.** Fig. 2 shows an example of a PV system scaling in Maricopa county, Arizona, US done for the year 2010. The county is host to the state capital, Phoenix. It contains 71,642 census data points, with altogether 3,824,191 inhabitants in 2010. Dimensioning the PV capacity such that it produces annually as much electricity as is used for cooling results in a peak PV capacity of approximately 5.9 GW. Fig. 2a shows the simulation results over the whole year and describes the seasonal similarities and differences of cooling demand and PV electricity production. Both are higher during the summer months (June – August) than during the winter months (December – February), but the two profiles are not perfectly aligned. Overall, the seasonal differences are greater for the cooling demand, than for PV production. Furthermore, in the spring, solar insolation begins to increase faster than temperature. After mid-Summer, the temperatures keep increasing, which increase cooling demand, but hamper PV module performance, while simultaneously the amount of sunlight per day begins to decrease.

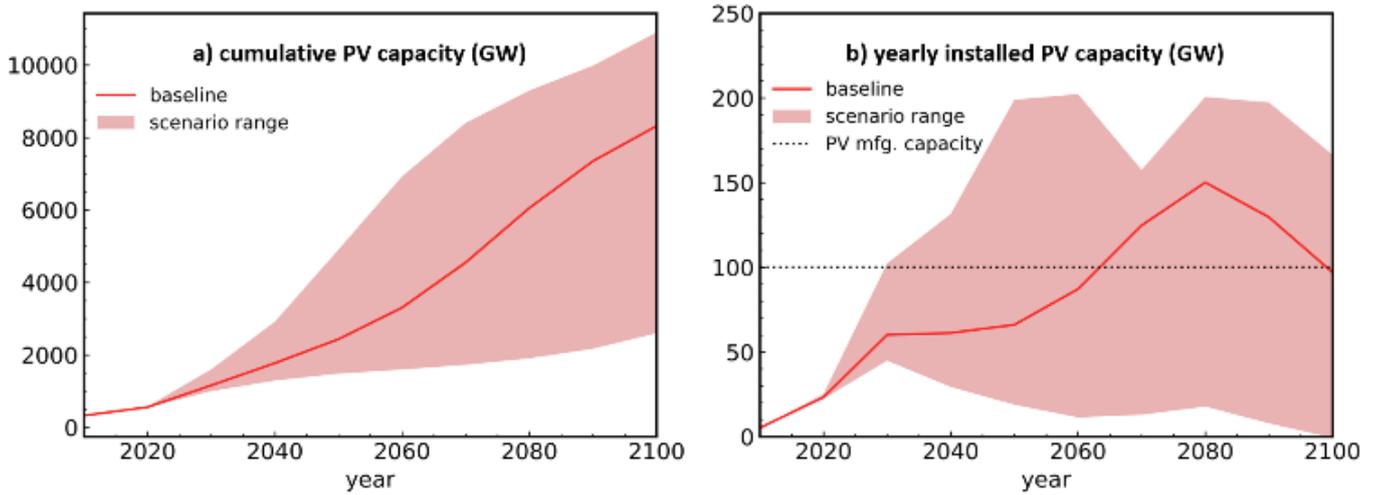

**Fig. 3** Cumulative (a) and yearly installed (b) PV capacity sustained by the cooling sector in the baseline scenario (red solid line) and the range of values within the scenarios explored (red shaded area). Also shown in (b) is the current PV manufacturing capacity (black dotted line).

Fig. 2 (b) highlights the differences and similarities in the diurnal variance of cooling demand and PV production, showing both profiles for one week at the beginning of May. Again, qualitatively, the variance is similar: both profiles are higher during the day and lower during the night. However, due to thermal inertia of the Earth, the temperature (cooling demand) lags behind solar insolation (PV production), causing PV overproduction during noon and underproduction during the evening.

Despite the different dynamics of cooling demand and PV production, the PV system can meet 55.5% of the cooling electricity demand on an hourly basis. Referring to Fig. 2, this means that over the whole year the "cool by PV" area marked by green is equal to 55.5% of the area under the blue "cooling" curve. In addition to energy storage, the temporal match of AC demand and PV production can be improved by orienting the PV array towards west[25,26,55-57] or using single- or double- axis tracking[54]. Additionally, buildings can possibly be pre-cooled during the sunniest hours of the days, so that they remain cool during the evening[58].

**Thermal storage model**

To demonstrate the potential of thermal storage to mitigate the hourly mismatch of cooling demand and PV electricity production, we investigate a scenario where every household with a cooling device is also equipped with a small-scale thermal storage. More specifically, we simulate a 1.0 m³ water-based latent heat storage, or a cube of ice and water which freezes as it is cooled and melts as it cools the residential building. The latent heat density of water is 334 kJ / kg. The storage is modelled to be at a constant temperature $T_{\text{storage}} = 0°C$. The storage operation is rule-based: it is cooled when there is PV overproduction, or until completely frozen, and it is melted when there is PV underproduction, or until it is completely liquid. The $COP$ for cooling the storage is calculated similarly as when cooling the building itself (see Eq. 6):

$$COP_{storage}(t) = \eta_{Ca} \times \frac{T(t) - \delta T_1}{T_{storage} + \delta T_2 - (T(t) - \delta T_1)}. \quad (9)$$

The storage walls are modelled to have an insulation $U$ value of 0.3 W/m²/K, which corresponds to an insulation provided by steel walls and 10 cm polyurethane insulation[30]. Cooling the building from the storage is expected to be lossless with $COP = 1$. The surroundings of the thermal storage are assumed to be at $T_{\text{base}}$.

## Results

**Projected Global Photovoltaic Capacity Sustained by Residential Cooling Sector**

Fig. 3 (a) depicts the projected PV generation capacity that the cooling sector could sustain in the baseline scenario (solid red line) and the range of scenarios explored (red shaded area). Already in 2018, the cooling sector could sustain approximately 540 GW of PV, more than the 402 GW installed PV production capacity at the end of 2017[59]. While the exact growth rate depends on the scenario, the PV capacity increases significantly in each explored scenario, reaching a capacity of 8.3 TW in the baseline scenario and a range of 2.6 TW-10.9 TW, an increase of a factor of 5-20, in the scenario range.

To elucidate the high growth speed of the added PV capacity, Fig. 3 (b) plots the projected yearly added PV capacity in the baseline scenario and the scenario range. The calculated yearly added PV capacity is between 20-200 GW throughout most of the century, which is at least 20%, and at most double the current PV manufacturing capacity which is approximately 100 GW/year[60]. This means that the residential cooling sector could alone demand a significant fraction of the global solar panel supply throughout the century.

We note that the empirical data on air-conditioner availability[42-47] used in this study was gathered before PV became cost-competitive with traditional energy sources. If PV continues its cost-reduction trends, it is plausible that cheap PV will enable cooling to be powered more cheaply, and thus be adopted at lower income levels and what is predicted by Eq. (4). Powering residential cooling via PV is further supported by the fact that self-consumption of PV is often incentivized for PV households[61] and PV systems tend to have decreasing unit prices as a function of system size (larger PV systems are cheaper per unit power than smaller systems)[62]. These facts could accelerate cooling demand, and PV capacity with it, faster than estimated here.

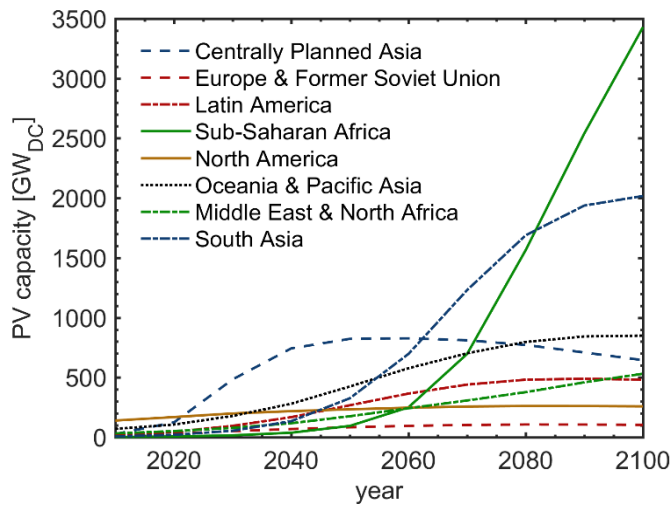

**Fig. 4** Regional assessment of projected PV capacity sustained by the residential cooling sector.

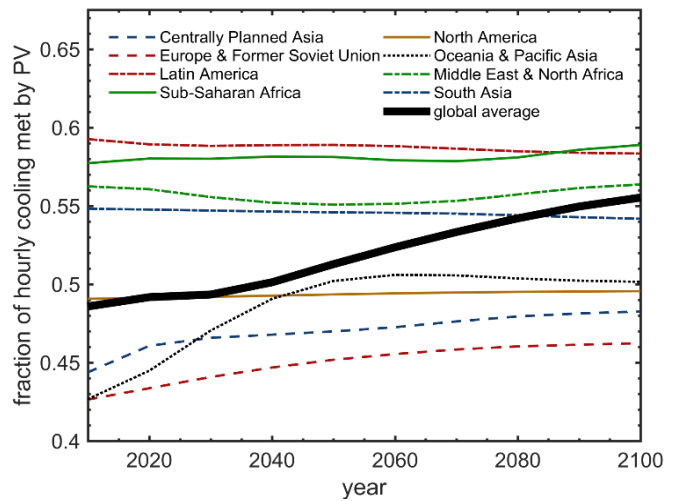

**Fig. 5** Fraction of cooling electricity consumption provided by photovoltaic electricity consumption on an hourly basis globally (thick black line) and regionally. This fraction is depicted in Fig. 2 b) by the quotient of the area under the blue curve and the green area. This figure assumes no storage greater than one hour, i.e., describes direct PV production to cooling power consumption.

**Projected Photovoltaic Capacity by Region**

Fig. 4 depicts a regional assessment of projected PV capacity in our baseline scenario. We divide the world into eight sectors, based on the regions used in the MESSAGE modelling framework developed by the International Institute for Applied Systems Analysis[63]. A list of countries and territories grouped in each region is listed Table S1. Today, the largest potential market for residential cooling and PV is in North America, mainly due to high income levels ($GDP/cap$ = \$US 45 330 in 2015) and reasonably warm climate ($CDD_a = 710$), particularly in the Southern US. The total PV generation capacity that residential cooling could sustain in North America is approximately 155 GW at the end of 2015. As incomes and population rise in other parts of the world, other regions become dominant.

In the baseline scenario, the first region with more cooling-sustained PV than North America is Centrally Planned Asia (China, Mongolia, Laos, Cambodia, Vietnam), which overtakes North America in 2021, as $GDP/cap$ in the region surpasses \$US 10 000. The potential capacity in this region grows and saturates to around 830 GW by 2050, after which increased demand per capita is offset by a decreasing population in the region and increased energy efficiency of cooling devices. Between 2020 to 2040, all other regions experience modest growth, with Oceania & Pacific Asia growing the fastest.

During the second half of the century, two regions grow significantly larger than others, mainly due to their high population: first South Asia, with a population of approximately 2 billion people, could sustain a 1 TW capacity by 2065, and ultimately over 2 TW by the end of the century. Lastly, Sub-Saharan Africa is the last region to reach income levels commensurate with significant air-conditioner adoption, reaching a $GDP/cap$ of \$US 10 000 around 2055. Due to its large population (2.35 billion by the end of the century) and hot climate ($CDD_a = 2\,670$ by 2100), it presents the largest potential region for additional PV capacity, with over 3 TW of potential PV capacity by the end of the century.

Slightly smaller, but still significant sectors include Middle East & North Africa, as well as Latin America. They exhibit similar potential throughout the century, reaching potential PV capacities of approximately 532 GW (Middle East & North Africa) and 482 GW (Latin America) at the end of the century.

Europe & Former Soviet Union, on the other hand, present a modest potential market compared to other sectors, mostly due to their Northern location, and hence, low cooling demand ($CDD_a = 350$ in 2100). The region reaches approximately 104 GW by the end of the century.

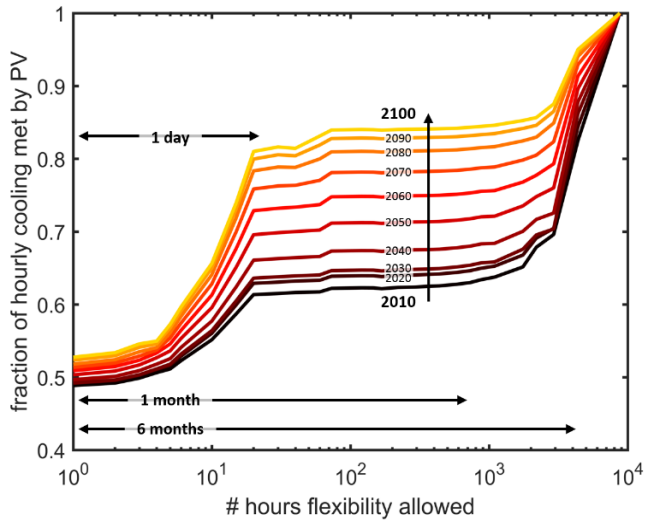

**Fig. 6** Different time scales of storage needed to fully power cooling by PV during the 21st century.

**Fraction of cooling electricity consumption provided by PV on an hourly basis**

Although the cooling electricity usage and PV production is matched over a year, not all cooling demand is directly satisfied via PV on an hourly basis due to the different seasonal and diurnal dynamics of cooling and PV (cf. Fig. 2). Fig. 5 presents the fraction of cooling electricity provided by PV on an hourly basis, averaged globally (thick black line) as well as regionally. As shown in Fig. 2, generally the seasonal differences in cooling demand are larger than those in PV energy yield. Because closer to the equator the seasonal differences are smaller, generally regions closer to the equator exhibit stronger synergy of cooling and PV, and hence, a larger fraction of the cooling electricity demand can be met on an hourly basis.

The two regions closest to the equator are Latin America and Sub-Saharan Africa, with population weighted geographical centres around 2°S and 1°N, respectively. In these regions, PV can directly meet as much as 57-59% of the cooling demand. Middle East & North Africa and South Asia are located slightly due north at 28°N and 24°N, respectively, with still quite high synergy with 54-56% of cooling met on an hourly basis. Of the remaining regions, Centrally Planned Asia, North America and Europe & Former Soviet Union, are located further north at 31°N, 38°N and 48°N, respectively, with consequently less cooling directly powered by PV. Overall, the average fraction within a region stays relatively constant, and the relative order of different regions remains similar. The one exception is Oceania & Pacific Asia where the synergy improves during the century. The improvement occurs because in the beginning of the century, most of the cooling consumption occurs in Pacific OECD countries, such as Australia and Japan, which are further from the equator than countries like Indonesia and Philippines, which experience greater cooling consumption growth towards the middle and latter half of the century.

The global average fraction of cooling sustained by PV increases significantly throughout the century, from approximately 49% to 56%. The change in the global average is mostly caused by global socioeconomic changes: as hot countries in Asia and Africa use more cooling, their more suitable climate for combining cooling & PV begins to dominate the global average.

Importantly, we note that if one would power a constant load (not cooling) with PV, one could only power this load for approximately 40% on an hourly basis. This implies that when combined with PV, the variability of cooling loads is an advantage, rather than a disadvantage.

**Storage analysis**

One way to increase the fraction of cooling powered by PV is via energy storage. It is interesting to investigate the different time-scales of the mismatch in cooling demand and PV energy yield. For example, diurnal mismatch is significantly easier to mitigate than seasonal mismatch, because all energy storages dissipate over time, and the total amount of energy to be stored tends to be larger with seasonal differences than diurnal differences (cf. Fig. 2). To investigate the different time-scales of the mismatch on a global scale, we split the year into separate increments of $n$ hour, where $n \in [1, N_h]$, and allowed PV overproduction to freely meet cooling demand within each increment. While this analysis does not directly compare to any physical energy storage, it is a straightforward way to elucidate the different time-scales of storage needed.

Fig. 6 shows the updated fraction of hourly cooling that was met with PV via this exercise for every year. We see that majority of the flexibility needed is either short-term (<1 day) or long-term (>1 month), with minimal additional benefit gained by potential flexibility between 1-30 days. Secondly, we see that the benefit of short-term storage increases greatly over time, as cooling demand grows in locations where the mismatch is mostly diurnal, rather than seasonal. In 2010, short-term flexibility could increase the fraction of hourly cooling met by PV from 49% to 61%, but in 2100, the increase would be from 53% to as much as 81%.

This suggest that over time, storage solutions will become more and more beneficial to residential buildings with cooling systems. To demonstrate this, we implemented the storage model described in Section 2.5 for every location and every year. The results are shown in Fig. 7. The increased benefit of storage as a function of time can be seen by the increasing fraction of cooling that the 1 m³ thermal ice storage helps meet. In 2010, the fraction is 6 %-points but in 2100, it is 15 %-points, or 2.5 times more.

**Sensitivity analysis: Impact of socioeconomic and global warming scenarios**

Fig. 8 elucidates the impact of the specific climate or socioeconomic scenario chosen. Specifically, Fig. 8 modifies the baseline scenario (2.5°C, SSP2) by either assuming less (1.8°C) or more (3.8°C) global warming by the end of the century, or a socioeconomic scenario with faster (SSP5) or slower (SSP3) development. In the near-future, the socioeconomic scenario plays a larger role, with 1 TW PV capacity reached between 2024 and 2030. Toward the end of the century, cooling demand begins to saturate in the fast development scenario, and it does not significantly differ from the baseline scenario anymore. The slow socioeconomic scenario results in significantly lower PV capacity, mostly because income levels in the developing world, such as South Asia and Sub-Saharan Africa, remain modest.

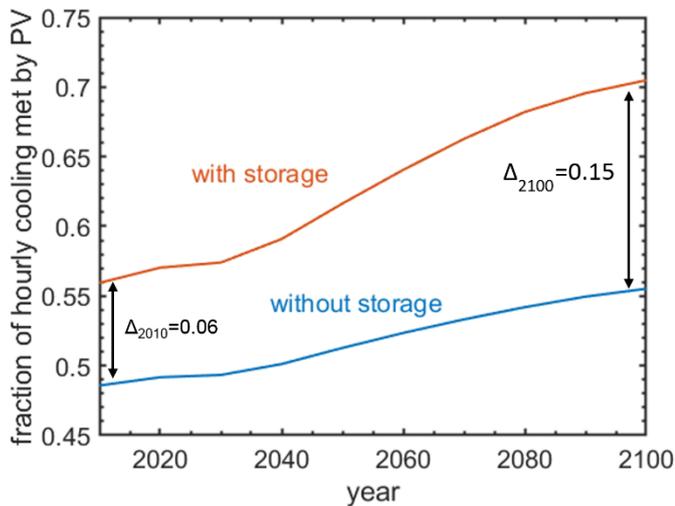

**Fig. 7** Fraction of hourly cooling directly met by PV without any storage longer than one hour (blue line) and assuming every household with a cooling device also has a 1 m³ thermal ice storage described in the Thermal Storage Model section (orange line).

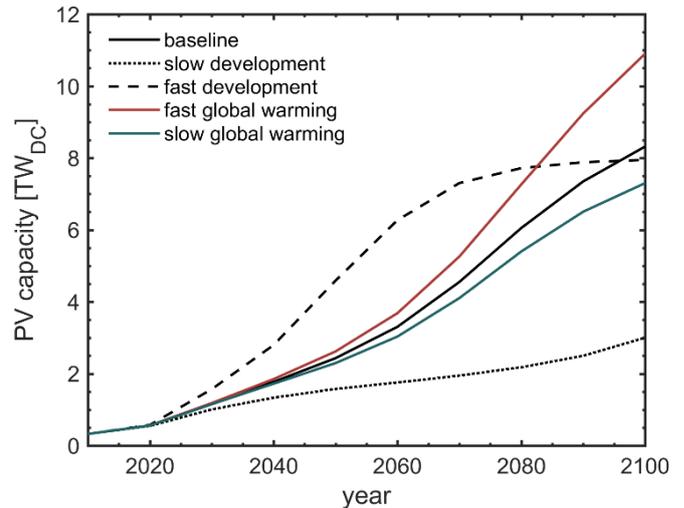

**Fig. 8** Dependence of the projected PV capacity on the climate and socioeconomic scenario.

Interestingly, the fast socioeconomic development scenario saturates at a level slightly below the baseline at the end of the century. This is because socioeconomic development is coupled with lower fertility rates, and hence the population at the end of the century is considerably lower in the fast development scenario (7.4 billion) as in the baseline (9.0 billion).

The global warming scenario, on the other hand, begins to play a role beyond 2050, and can decrease the end of the century PV capacity by 1 TW or increase it by 2 TW. Regardless of the scenarios chosen, the residential cooling sector can sustain significant amounts of PV capacity, with yearly additions commensurate with at least several tens of per cents of the current PV manufacturing capacity. In terms of the hourly dynamics of cooling demand and PV energy yield, the scenario choice is of modest importance, and PV can meet approximately 50 % of the cooling demand on an hourly basis regardless of scenario.

## Conclusions

Cooling demand is the fastest-growing form of energy demand in buildings and solar photovoltaics is the fastest growing form of energy production. While both are intrinsically variable and thus a challenge to integrate into broader energy systems, their variance is fortunately ultimately driven by the same source: the sun.

Here, we showed that if all electricity used by residential cooling devices globally was supplied via PV, the residential cooling sector could sustain approximately 540 GW of PV generation capacity, on par with the global installed capacity today (402 GW at the end of 2017[60]). Additionally, we showed that due to tropical countries gaining more wealth and thus access to cooling and global warming advancing, the cooling devices could sustain another 20-200 GW of PV every year throughout most of the 21st century, depending on the chosen socioeconomic and climate change scenario, despite air-conditioners growing more efficient over time. This implies that the residential cooling sector could act as a significant growth driver for the global PV industry throughout the century, with the same likely holding true for the commercial sector as well.

Additionally, we modelled the hourly dynamics of both PV production and cooling demand, and found that without any storage, PV could directly power approximately 50% of the cooling demand, with the fraction increasing from 49% to 55% in our baseline scenario during the 21st century. The increase in synergy is due to the geographic shift in cooling demand closer to the equator, where seasonal differences in weather are smaller, and the temporal dynamics of PV and cooling better aligned. This trend remains regardless of scenario choice.

With seasonal differences growing smaller, also the impact of small-scale distributed storage grows larger. We simulated a scenario in which every household with a cooling device was equipped with a 1 m³ water-based latent thermal storage. In the beginning of the century, the storage increases the fraction of cooling demand powered by PV from 49% to 56%, whereas in the end of the century, the fraction grows from 55% up to 70%.

These results demonstrate that a clear majority of the rapidly increasing cooling demand can be met sustainably with PV and small-scale distributed storage and that the intersection of cooling and PV is and continues to be a high-impact target for continued research, business and policy investments.

## Outlook

The approach used here to investigate the synergy of cooling and PV could likely be expanded into other sectors with significant growth expected and high intrinsic synergy with PV electricity production.

For example, to avoid catastrophic climate change, nearly the entire transportation sector needs to switch from fossil fuels to electric batteries or synthetic fuels by the end of the 21st century. In 2012, the global transport sector consumed approximately 30 000 TWh of primary energy[64]. Since both batteries and synthetic fuels can store PV electricity, it is plausible that PV could provide a significant fraction of this energy. To produce a corresponding amount of PV electricity with a capacity factor of 20% would require as much as 17 TW of PV capacity.

Water desalination is another interesting sector for added PV capacity as the highest demand for desalination is in hot and dry climates with high PV capacity factors. Its demand will also likely increase as hot and dry countries in for example the Middle East and Africa gain wealth[65]. Desalinated water is also relatively easy to store, which makes its production highly flexible[66]. Other examples include raw material manufacturing plants such as those of aluminum or pulp[67]. Since these plants can rapidly ramp their production up and down, their electricity usage can match PV electricity generation. Performing a similar analysis for these sectors as we did here for cooling would help clarify when, where and to what extent the global PV industry can grow.

## Conflicts of interest

There are no conflicts to declare.

## Acknowledgements

H. S. Laine acknowledges the Fulbright Technology Industries of Finland grant and the support of the Finnish Cultural Foundation and Tiina and Antti Herlin Foundation. Ashley Morishige and Gregory Wilson are acknowledged for fruitful discussions.

**Electronic Supplementary Material for:**

**Meeting Global Cooling Demand with Photovoltaics during the 21$^{st}$ Century**

Hannu S. Laine[a,b], Jyri Salpakari[c], Erin E. Looney[a], Hele Savin[b], Ian Marius Peters[a], and Tonio Buonassisi[a]

[a]Massachusetts Institute of Technology, Cambridge, MA 02139, USA.

[b]Department of Electronics and Nanoengineering, Aalto University, 02150 Espoo, Finland.

[c]New Energy Technologies Group, Department of Applied Physics, Aalto University, 02150 Espoo, Finland.

Table S1: Countries and territories included in each region.

| Region | Countries included in the region |
|---|---|
| Centrally Planned Asia | Cambodia, China (incl. Hong Kong & Macao), Lao People's Democratic Republic, Mongolia, Vietnam |
| Europe & Former Soviet Union | Albania, Armenia, Austria, Azerbaijan, Belgium, Bulgaria, Bosnia and Herzegovina, Belarus, Croatia, Cyprus, Czech Republic, Denmark, Estonia, Finland, France, Georgia, Germany, Greece, Hungary, Iceland, Ireland, Italy, Kazakhstan, Kosovo, Kyrgyzstan, Latvia, Lithuania, Luxembourg, Malta, Moldova, Montenegro, Netherlands, Norway, Poland, Portugal, Romania, Russia, Serbia, Slovakia, Slovenia, Spain, Sweden, Switzerland, Tajikistan, Turkey, Turkmenistan, The former Yugoslav Republic of Macedonia, Ukraine, United Kingdom of Great Britain and Northern Ireland, Uzbekistan |
| Latin America | Argentina, Bahamas, Belize, Bolivia, Brazil, Barbados, Chile, Colombia, Costa Rica, Cuba, Dominican Republic, Ecuador, El Salvador, Guatemala, Guyana, Honduras, Haiti, Jamaica, Mexico, Nicaragua, Panama, Peru, Paraguay, Suriname, Trinidad and Tobago, Uruguay, Venezuela |
| Sub-Saharan Africa | Angola, Benin, Botswana, Burkina Faso, Burundi, Cameroon, Cape Verde, Central African Republic, Chad, Comoros, Congo, Côte d'Ivoire, Democratic Republic of the Congo, Djibouti, Eritrea, Ethiopia, Gabon, Gambia, Ghana, Guinea, Guinea-Bissau, Equatorial Guinea, Kenia, Lesotho, Liberia, Madagascar, Malawi, Mali, Mauritania, Mauritius, Mozambique, Namibia, Niger, Nigeria, Rwanda, Senegal, Sierra Leone, Somalia, South Africa, Swaziland, Togo, Uganda, United Republic of Tanzania, Western Sahara, Zambia, Zimbabwe |
| North America | Canada, Puerto Rico, United States |
| Oceania & Pacific Asia | Australia, Brunei Darussalam, Fiji, French Polynesia, Indonesia, Japan, Republic of Korea, Malaysia, Myanmar, New Caledonia, New Zealand, Papua New Guinea, Philippines, Singapore, Solomon Islands, Taiwan, Thailand, Timor-Leste, Vietnam, Western Samoa |
| Middle East & North Africa | Algeria, Bahrain, Egypt, Iran, Iraq, Jordan, Israel, Kuwait, Lebanon, Libya, Morocco, Oman, Palestine, Qatar, Saudi Arabia, South Sudan, Sudan, Syria, Tunisia, United Arab Emirates, Yemen |
| South Asia | Afghanistan, Bangladesh, Bhutan, India, Maldives, Nepal, Pakistan, Sri Lanka |